\documentstyle{article}

\input tcilatex
\begin{document}

\begin{center}
\textbf{TRAPPING, COMPRESSION AND ACCELERATION OF AN ELECTRON BUNCH IN THE
NONLINEAR LASER WAKEFIELD}

Arsen G. Khachatryan

\textit{Yerevan Physics Institute, Alikhanian Brothers Street 2, }

\textit{Yerevan 375036, Armenia}
\end{center}

A scheme of laser wakefield acceleration, when a relatively rare and long
bunch of non-relativistic or weakly-relativistic electrons is initially in
front of the laser pulse, is suggested and considered. The motion of test
electrons is studied both in the one-dimensional case (1D wakefield) and in
the three-dimensional laser wakefield excited in a plasma channel. It is
shown that the bunch is trapped, effectively compressed both in longitudinal
and transverse directions and accelerated to ultra-relativistic energies in
the region of first accelerating maximum of the wakefield. The accelerated
bunch has sizes much less than the plasma wavelength and relatively small
energy spread.

\ 


\ 

\begin{center}
\textbf{I. INTRODUCTION}
\end{center}

The rapid progress in the technology of high-intensity lasers, based on the
chirped-pulse amplification (CPA) [1], opens new opportunities for the use
of lasers in many branches of science and industry. Relatively inexpensive
tabletop terawatt lasers (so-called T$^3$-lasers) become a qualitatively new
tool in physical researches and now are available at many laboratories over
the world. Last years CPA technique permitted the production of
subpicosecond laser pulses of multiterawatt power with peak intensities
exceeding 10$^{20}$W/cm$^2$ [2]. With intensities as such we practically
have to do with a new interaction range of laser radiation with matter,
where the role of the nonlinear effects is often essential. In this intense
laser field the matter is usually transformed to plasma and free electrons
oscillate with relativistic quivering energy. Presently, the interactions of
high-power laser radiation with plasma are actively investigated in
connection with different applications: the excitation of strong plasma wake
waves for focusing and acceleration of charged bunches [3]; generation of
radiation at harmonics of carrier laser frequency [4]; X-ray sources [5];
laser inertial fusion [6] etc.

The laser wakefield, generated in plasma by the short (with the length $%
\approx \lambda _p$/2, where $\lambda _p$ is the plasma wavelength) intense
laser pulse provides the acceleration gradient up to tens GeV/m (laser
wakefield acceleration, LWFA [7,8]), that is three orders of magnitude
higher than that achieved in conventional accelerators. The main aim of
experimental and theoretical works, that are presently in progress, is the
construction of compact and relatively inexpensive accelerators of charged
particles for applications in physics research, medicine and hi-tech
industry. However, some challenges remain on this way, one of the main of
those is the problem of electron bunch injection.

The wake wavelength in the LWFA is $\lambda _p\approx 2c\tau _L$ [3] (here $%
\tau _L$ is the laser pulse duration) and makes up tens or hundreds
micrometers for typical plasma densities $n_p\sim 10^{16}-10^{19}$ cm$^{-3}$%
. To obtain high quality relativistic electron bunch accelerated by the wake
wave, it is necessary to inject short (with the length $L\ll \lambda _p$),
enough dense relativistic electron bunch in the accelerating phase of the
wake wave with femtosecond synchronization, that is difficult technical
problem (see e.g. Ref. [9]). The injection schemes proposed for the standard
LWFA (the LIPA [9], the colliding laser pulses [10] and the LILAC [11]
schemes) are aimed at the generation of such a short relativistic bunch.

The diffraction broadening leads to rapid decrease of the intense laser
pulse amplitude with the characteristic length $Z_R=\pi r_0^2/\lambda _L$
(here $Z_R$ is the Rayleigh length, $r_0$ is the focal spot size of the
pulse, and $\lambda _L$ is the laser wavelength) that is typically in order
of a millimeter. To prevent diffraction the plasma channel, with minimum
density at the axis, proposed to guide the laser pulse in the LWFA [12]. The
amplitude of the accelerating component of the wake wave, generated in the
plasma channel, decreases as the distance from the laser pulse increases
[13,14]. Besides, the change of the plasma wavelength $\lambda _p$ in the
transverse direction [$\lambda _p\sim n_p^{-1/2}(r)$, where $n_p$ is the
density of electrons in the plasma channel], leads to undesirable wave front
curving; this effect becomes stronger as the distance from the pulse
increases. The effect of the wave front curving in the channel, in the case
of a strong laser pulse ($a_0=eE_m/m_ec\omega _L\geq 1$, where $e$ and $m_e$
are the absolute charge and mass of the electron, $E_m$ is the maximum
amplitude of laser field, $c$ is the speed of light in vacuum, and $\omega
_L $ is the laser frequency) is amplified due to the nonlinear change of the
wake wavelength in transverse direction [15,16]. Thus, for regular
acceleration of a charged bunch in the wake wave, the most preferable is the
region of the first maximum of accelerating field behind the laser pulse.

To avoid the aforesaid difficulties in LWFA, we suggest and study in this
work a scheme of trapping, compression and acceleration of a
non-relativistic or weakly-relativistic electron bunch in the laser
wakefield, when the bunch is initially in front of the laser pulse. The
initial bunch density can be much less than that required for the
accelerating bunch and the bunch sizes - in order or more than the plasma
wavelength, i.e. much more than require other methods of injection [9-11].
Our investigations take into account both the pulse ponderomotive force and
the wakefield. It is shown that the bunch is trapped, effectively compressed
both in longitudinal and transverse directions and accelerated to
ultra-relativistic energies in the region of first accelerating maximum of
the wakefield. The accelerated bunch has sizes much less than the plasma
wavelength and enough good quality.

\begin{center}
\textbf{II. THE CASE OF WIDE LASER PULSE}
\end{center}

At first we neglect the transverse variation of the laser pulse amplitude
and consider the case of one-dimensional laser wakefield excited by a wide
bunch in uniform plasma. This allows to study the longitudinal dynamics of
the bunch electrons in more details.

\begin{center}
\textbf{A. Basic equations and correlations}
\end{center}

One-dimensional steady wakefield excited by the linearly-polarized laser
pulses are described by the following equation (see, e.g. Ref. [3])

\begin{equation}
\frac{d^2\Phi }{d\xi ^2}+\beta _g\gamma _g^2\left\{ 1-\beta _g\frac{\Phi
/(1+a^2/2)^{1/2}}{[\Phi ^2/(1+a^2/2)-\gamma _g^{-2}]^{1/2}}\right\} =0, 
\tag{1}
\end{equation}
where $\Phi =1+e\varphi /m_ec^2$ is the dimensionless potential of the
plasma wakefield, $a=eE_0(\xi )/m_ec\omega _L$, $E_0$ is the electric field
amplitude of the laser pulse, $\xi =k_p(z-v_gt)$, $k_p=\omega _p/v_g$, $%
\omega _p=(4\pi n_pe^2/m_e)^{1/2}$ is the plasma frequency, $v_g$ is the
group velocity of the laser pulse which is equal to the phase velocity of
the wake wave, $\beta _g=v_g/c$, $\gamma _g^{}=(1-\beta _g^2)^{-1/2}$ is the
relativistic factor, which, in the case $\gamma _g^{}>>1$, is nearly equal
to $\omega _L/\omega _p$. The electric field of exited wakefield, normalized
to the non-relativistic wave-breaking field $E_{WB}=m_ev_g\omega _p/e$, can
be obtained from equation $E_z=-(1/\beta _g)^2d\Phi /d\xi $. The equation of
motion of the test electron in the wakefield and in the field of laser pulse
is (see e.g. Ref. [17]):

\begin{equation}
\frac{dp}{d\tau }=-\frac 1{4\beta _g\gamma }\frac{da^2}{d\xi }-\beta _gE_z, 
\tag{2}
\end{equation}
Here the first term on the right-hand side is the relativistic ponderomotive
force averaged over the fast laser oscillations, and the second one
corresponds to the plasma wakefield excited by the laser pulse, $\beta
_{}=v_{}/c$, $p=\beta \gamma $ and $\gamma
=(1+p^2+a^2/2)^{1/2}=[(1+a^2/2)/(1-\beta _{}^2)]^{1/2}$ are the normalized
longitudinal velocity and momentum and the relativistic factor of the test
electron (transverse velocity is zero in this Section), $\tau =\omega _pt$.
Multiplying Eq. (2) by $\beta $, after some simple transformations, one can
obtain the following integral of motion (see also Refs. [14,18])

\begin{equation}
\gamma -\beta _gp-\Phi =const.  \tag{3}
\end{equation}
Let us consider an electron that is initially ahead of the laser pulse at
some point $\xi _0$ where $\Phi =1$ and $a=0$. If the electron has initial
momentum $p_0<\beta _g\gamma _g$, it will be overtaken by the laser pulse
and can be trapped at some point $\xi _r$ inside the pulse or in the wake
and accelerated. At the trapping point (or, in other words, at the point of
reflection) the velocity of the electron become equal to $v_g$. Then from
expression (3) we have:

\begin{equation}
S\equiv (1+a_r^2/2)^{1/2}/\gamma _g-(\Phi _r-1)=(1+p^2)^{1/2}-\beta _gp. 
\tag{4}
\end{equation}
In Eq. (4) $a_r$ and $\Phi _r$ are the amplitude of the laser pulse and the
wakefield potential at the reflection point $\xi _r$. From Eq. (4) it follows

\begin{equation}
p=\gamma _g[\beta _g\gamma _gS\pm (\gamma _g^2S^2-1)_{}^{1/2}].  \tag{5}
\end{equation}
The minus sign in (5) corresponds to the initial momentum $p_0$ of an
electron which has momentum $\beta _g\gamma _g$ at the point $\xi _r$ and
the plus sign corresponds to the momentum of free electron which initially
was at the point $\xi _r$. Expression (5) describes both trapped and passing
particles. In the wake, electrons can be trapped only in the region where $%
E_z\leq 0$. Equation of motion (2) can be rewritten in the form

\begin{equation}
\frac{d^2\xi }{d\tau ^2}+\frac{(1-\beta _g\beta )}{4\beta _g^2\gamma ^2}%
\frac{da^2}{d\xi }+\frac{(1-\beta _{}^2)}\gamma E_z=0,  \tag{6}
\end{equation}
where $\xi $ is the coordinate of a test electron in the frame commoving
with the laser pulse. The dimensionless velocity of the electron one can
obtain from expression $\beta =\beta _g(1+d\xi /d\tau )$.

\begin{center}
\textbf{B. Numerical results}
\end{center}

Eqs. (1) and (6) were solved numerically for the Gaussian laser pulse,

\[
a=a_0\exp [-(\xi -\xi _c)^2/\sigma _z^2]. 
\]
In Fig. 1 a laser pulse with $a_0=2$ and nonlinear wakefield excited by it
are presented (here and below in numerical calculations $\sigma _z=2$, $\xi
_c=3$ and $\gamma _g=10$). The amplitude of the wake wave is essentially
less than one-dimensional relativistic wave-breaking field $%
E_{rel}=[2(1-\gamma _g)]^{1/2}/\beta _g\approx 4.26$ [3,19]. Figure 2 shows
the dependence of initial electron momentum $p_0$ on the trapping point near
the first accelerating maximum in the wake wave. The minimum value of the
initial momentum $p_{\min }$ corresponds to the trapping point where the
potential is at the minimum and $E_z=0$. Curves 1 and 2 in Fig. 2 reach
their minimums at different points, that is the consequence of the nonlinear
increase of wake wavelength with the amplitude (this dependence can be found
in Ref. [20]). The curves were obtained numerically and coincide with the
expression (5) for the trapped particles. Fig. 3 shows the dependence of the
value of $p_{\min }$ and wake wave amplitude $E_{z,\max }$ on $a_0$. One can
see that a laser pulse with $a_0\sim 1$ (that corresponds to the peak
intensity of the pulse $I_{\max }\sim 10^{18}W/cm^2$ for $\lambda _L=1\mu m$%
, and $I_{\max }\sim 10^{16}W/cm^2$ when $\lambda _L=10\mu m$) provides
trapping of initially non-relativistic or weakly-relativistic electrons in
the wake wave. For example, $p_{\min }\approx 0.4$ for the wakefield
presented in Fig. 1. Electrons with $p_0<p_{\min }$ can not be trapped in
the wake wave and may be detected behind the wave. This circumstance can
help to determine the wake wave amplitude in experiments. Our numerical
calculations have witnessed that electrons with $p_0<\beta _g\gamma _g$ can
not be trapped in the region occupied by the laser pulse because of the
decelerating wakefield; only electrons with $p_0\approx \beta _g\gamma _g$
are trapped in the head of the pulse (where $E_z\approx 0$) due to the
ponderomotive force. This confirms with the results of Ref. [18].

Figure 4 shows the behavior of electrons of mono-energetic non-relativistic (%
$p_0=0.5,$ $\gamma _0=(1+p_0^2)^{1/2}\approx 1.12$) bunch in the wakefield
presented in Fig. 1. Initial dimensionless bunch length is $L_0=5$, that
roughly corresponds to the linear plasma wavelength $\lambda _p$. When $\tau
=50$, the trapped bunch length is $L\approx 0.027$ and $L\approx 0.04$ when $%
\tau =100$, that is two orders of magnitude less than the initial bunch
length. The absolute energy spread $\Delta \gamma $ in the accelerating
bunch increases insignificantly with time, but the relative energy spread $%
\varepsilon =\Delta \gamma /\gamma $ falls due to growing $\gamma $; for
example $\varepsilon \approx 0.26$ at $\tau =50$, and $\varepsilon \approx
0.14$ when $\tau =100$. The acceleration gradient in the considering case is
approximately equal to $2MeV/\lambda _p$. For example, when $\lambda
_p=100\mu m$ ($n_p\approx 10^{17}cm^{-3}$), the acceleration gradient is $%
20GeV/m$.

Figure 5 shows the motion of electrons with different initial momentums and
the same initial positions ($0.6\leq p_0\leq 1.2,$ $1.17\leq \gamma \leq
1.56,$ $\xi _0=0$) in the wakefield presented in Fig. 1. The trapped bunch
length is nearly 27 times less than the plasma wavelength $\lambda _p$. The
relative energy spread at $\tau =100$ is about $0.1$, that is much less than
that of initial electrons.

The dephasing length, for electrons with $p_{\min }\leq p_0\leq 1.2$, varies
in the range $630\leq L_d\leq 700$ (the grater values correspond to the
smaller initial momentums) that is comparable with the linear dephasing
length $\lambda _p\gamma _g^2$ [3], which, in our notations, corresponds to $%
L_d=2\pi \gamma _g^2=200\pi $. The maximum relativistic factor of
accelerated particles is in the range $350\leq \gamma _{\max }\leq 410$
(here again the greater values correspond to smaller $p_0$) that essentially
exceeds the linear value $2\gamma _g^2=200$ [3], but is an order of
magnitude less than the maximum nonlinear value $4\gamma _g^3=4000$ [20,21].

\begin{center}
\textbf{C. Energy spread in the accelerating bunch}
\end{center}

Energy spread in the trapped bunch depends on energy spread and length of
the initial bunch. The tail electrons of initial bunch are trapped earlier
and therefore, have greater energy during acceleration (see Fig. 4). Slower
particles also are trapped earlier (see Fig. 5). Let us suppose that
initially the bunch is at the head of laser pulse, so that $\xi =0$
corresponds to the bunch tail, and $\tau _{tr}(p_0)$ is the time necessary
to trap an electron which is initially at $\xi =0$; the initial electron
momentum is in the range $p_1\leq p_0\leq p_2$. Then, for energy spread in
the trapped bunch we can write $\Delta \gamma \sim \Delta \tau
_{tr}E_{z,\max }=[\tau _{tr}(p_2)-\tau _{tr}(p_1)+L_0/(1-v_2/v_g)]E_{z,\max
} $, where $\Delta \tau _{tr}$ is the time interval which is necessary to
trap the initial bunch. For the relative energy spread one has: $\varepsilon
\sim \Delta \tau _{tr}/(\tau -\Delta \tau _{tr})$. These estimates agree
well with the numerical results. One can see that the presence of fast
electrons (with $v_0\sim v_g\approx c$) in the initial bunch leads to an
undesirable increase in the energy spread.

The trapped bunch density can be found from expression $n_b(\tau )\approx
n_{b0}L_0/L(\tau )$, where $n_{b0}$ is the initial bunch density.

\begin{center}
\textbf{D. Wakefield generated by the accelerating bunch}
\end{center}

The trapped bunch also generates wakefield which can destroy the laser
wakefield and decrease the accelerating field. Because the accelerating
bunch is short ($L(\tau )\ll \lambda _p$) we can consider it as a plane
bunch and find the normalized amplitude of the wakefield excited by the
bunch from expression $E_{b,\max }=k_p(v_b/c)(N_b/n_p)$ [22], where $v_b$
and $N_b$ are velocity and the surface density of the bunch correspondingly.
This expression is valid both in linear and non-linear regimes. In our case $%
N_b=\delta n_{b0}L_0/k_p$, where $\delta \leq 1$ is the ratio of number of
trapped electrons to the total number of particles in the initial bunch, and
we have:

\begin{equation}
E_{b,\max }=\delta (v_b/c)(n_{b0}L_0/n_p).  \tag{7}
\end{equation}
The normalized amplitude of moderately nonlinear laser wake wave,
considering in this paper, is about unit. So, we can neglect the wakefield
generated by the bunch if $E_{b,\max }\ll 1$, or when

\[
n_{b0}\ll n_p(c/v_b)(1/\delta L_0). 
\]
For $n_p\sim 10^{16}-10^{18}cm^{-3}$ (that is typical for the LWFA
experiments [3]), $v_b\approx c$, $\delta \approx 1$ and the initial bunch
length in order of $\lambda _p$ ($L_0\sim 5-10$) this condition reads $%
n_{b0}<10^{14}-10^{16}cm^{-3}$. The density of accelerating bunch may be in
order of plasma density.

Thus, the one-dimensional analysis has showed the possibility of trapping,
essential compression and high-gradient acceleration of a low energy
electron bunch in moderately nonlinear laser wakefield.

\begin{center}
\textbf{III. TRAPPING, COMPRESSION AND ACCELERATION IN THE LASER WAKEFIELD
EXCITED IN PLASMA CHANNEL}
\end{center}

In this section we consider our scheme of LWFA for the case of laser
wakefield excited in a plasma channel and study the peculiarities of radial
motion of test electrons during trapping and acceleration.

\begin{center}
\textbf{A. Nonlinear laser wakefield excited in plasma channel}
\end{center}

As was mentioned in Introduction, the plasma channel is necessary to guide a
laser pulse. This allows to essentially increase the laser-plasma
interaction distance [12], that, in its turn, provides ultra-relativistic
acceleration in the wakefield [3]. Nonlinear axially-symmetrical laser
wakefields excited in a plasma channel are described by the following system
of equations [15]:

\begin{equation}
\beta \frac{\partial p_z}{\partial \xi }-\frac{\partial \gamma _e}{\partial
\xi }-\beta ^2E_z=0,  \tag{8.1}
\end{equation}

\begin{equation}
\beta \frac{\partial p_r}{\partial \xi }-\frac{\partial \gamma _e}{\partial r%
}-\beta ^2E_r=0,  \tag{8.2}
\end{equation}
\begin{equation}
-\frac{\partial H_\theta }{\partial \xi }+\beta \frac{\partial E_r}{\partial
\xi }+\beta _rN_e=0,  \tag{8.3}
\end{equation}
\begin{equation}
\nabla _{\bot }H_\theta +\beta \frac{\partial E_z}{\partial \xi }+\beta
_zN_e=0,  \tag{8.4}
\end{equation}
\begin{equation}
\beta \frac{\partial H_\theta }{\partial \xi }-\frac{\partial E_r}{\partial
\xi }+\frac{\partial E_z}{\partial r}=0,  \tag{8.5}
\end{equation}
\begin{equation}
N_e=N_p(r)-\nabla _{\bot }E_r-\frac{\partial E_z}{\partial \xi },  \tag{8.6}
\end{equation}
where $E_{z,r}$ and $H_\theta $ are longitudinal and radial components of
the electric field and azimuthal component of the magnetic field normalized
to the on-axis wave-breaking field $E_{WB}(r=0)=m_e\omega _p(r=0)v_g/e$, $%
p_{z,r}$ are the normalized components of plasma electron momentum, $\gamma
_e=(1+p_z^2+p_r^2+a^2/2)^{1/2}$ is the relativistic factor, $\beta _{z,\text{
}r}=p_{z,\text{ }r}$ $/\gamma _e$, $N_e=n_e(\xi ,r)/n_p(0)$ is the
normalized density of plasma electrons, $n_p(r)$ is unperturbed plasma
density in the channel, $N_p=n_p(r)/n_p(0)$, $\nabla _{\bot }=\partial
/\partial r+1/r$. The force acting on the relativistic electrons in the
wakefield is $\mathbf{F}(-eE_z,-e(E_r-\beta H_\theta ),0)$. According to
(8.5)

\begin{equation}
\frac{\partial E_z}{\partial r}=\frac{\partial (E_r-\beta H_\theta )}{%
\partial \xi }\equiv -\frac{\partial f_r}{\partial \xi }.  \tag{9}
\end{equation}
So, the field of forces $\mathbf{F}$ is potential because $\mathbf{%
\bigtriangledown }\times \mathbf{F}=0$ , and one can write $\mathbf{F}=%
\mathbf{\bigtriangledown }\Phi (\xi ,r)$, here $\Phi =1-\int_\xi ^0E_zd\xi $.

In this section we consider an axially-symmetric laser pulse which has
Gaussian profile both in longitudinal and radial directions:

\[
a(\xi ,r)=a_0\exp [-(\xi -\xi _c)^2/\sigma _z^2]\exp (-r^2/\sigma _r^2). 
\]
The laser pulse is guided in preformed plasma channel which has the
following unperturbed electron density:

\begin{equation}
N_p=\left( 1+\Delta \frac{r^2}{r_{ch}^2}\right) \exp \left( -b\frac{r^4}{%
r_{ch}^4}\right) ,  \tag{10}
\end{equation}
where $r_{ch}$, $\Delta $ and $b\ll 1$ are constant values. Such a density
profile is typical for plasma channels created in experiments [23]. Suppose
that the pulse is guiding without change in its radius $\sigma _r$. In this
case $\sigma _r=r_{ch}$ and $n_p(r_{ch})-n_p(0)=1/\pi r_er_{ch}^2$, where $%
r_e=e^2/m_ec^2\approx 2.8\times 10^{-13}cm$ is the classical electron radius
and all values are dimensional [12]. Then, in expression (10), $\Delta
=(2/\sigma _r\beta _g)^2$.

Equations (8.1)-(8.6) were solved numerically for the following parameters
of the problem: $a_0=2$, $\sigma _z=2$, $\sigma _r=5$ and $\gamma _g=10$. In
this case $\Delta \approx 0.16$, the value of $b$ was chosen to be 0.01. In
Fig. 7 we present the radial profile of unperturbed plasma density and the
radial behavior of laser pulse intensity, namely $exp(-2r^2/\sigma _r^2)$.
Fig. 8 shows the longitudinal electric field and the focusing field $%
f_r=\beta H_\theta -E_r$ of the wakefield excited. One can see that the wake
wavelength decreases as $r$ increases. This is caused by the radial increase
of unperturbed plasma density in the channel [3,13] and by the nonlinear
increase of wavelength with the wake wave amplitude which is at maximum on
the axis [15,16,24]. Fig. 7 shows also the nonlinear steepening of the
accelerating field like that takes place in one-dimensional wakefield (see
Fig. 1). Due to the dependence of the wavelength on $r$, the field in the
radial direction grows more chaotic as the distance from the laser pulse
increases. In fact, the oscillations of the plasma for different $r$ are
started behind the pulse with nearly equal phases but different wavelengths.
As $|\xi |$ increases, the change of phase in the transverse direction
becomes more and more marked. This leads to a curving of the phase front and
to oscillations in the transverse direction [15,16,24]. Such behavior of the
wakefield excited in a plasma channel leads to the transverse multistream
motion of plasma electrons in the wake and to the transverse wave-breaking
[25]. The radial dependence of longitudinal electric field and the focusing
force is shown in Fig. 8 for point $\xi =-10.9$ at which the on-axis
accelerating field reaches its maximum. We see that the wakefield changes
its sign and is steepened. For the ultra-relativistic acceleration of
electrons one needs to use a region in the wakefield where the conditions $%
E_z<0$ and $f_r<0$ are satisfied simultaneously. The radial steepening leads
to the radial restriction or the region suitable for acceleration. Near the
first accelerating maximum of the wakefield shown in Fig. 7, the suitable
region is $r<2.8$. As the distance from the laser pulse increases, the
suitable region becomes narrower, so that at some distance the wakefield is
highly irregular. Thus, the most preferable for electron acceleration is the
region of the first accelerating maximum in the wake.

\begin{center}
\textbf{B. Equation of motion of bunch electrons}
\end{center}

Three-dimensional vector equation of motion of bunch electrons is

\begin{equation}
\frac{d\mathbf{p}}{d\tau }=-\beta _g(\mathbf{E}+\mathbf{\beta }\times 
\mathbf{H})-\frac 1{4\beta _g\gamma }\bigtriangledown a^2.  \tag{11}
\end{equation}
Here all values are dimensionless, $\mathbf{\beta =v}/c=\mathbf{p}/\gamma $
is the normalized velocity, $\gamma =(1+\mathbf{p}%
^2+a^2/2)^{1/2}=[(1+a^2/2)/(1-\mathbf{\beta }^2)]^{1/2}$ is the relativistic
factor. For the momentum components, from Eq. (11) one has:

\begin{equation}
\frac{dp_r}{d\tau }=-\beta _g(E_r-\beta _zH_\theta )-\frac 1{4\beta _g\gamma
}\frac{\partial a^2}{\partial r},  \tag{12.1}
\end{equation}

\begin{equation}
\frac{dp_\theta }{d\tau }=0,  \tag{12.2}
\end{equation}

\begin{equation}
\frac{dp_z}{d\tau }=-\beta _g(E_z+\beta _rH_\theta )-\frac 1{4\beta _g\gamma
}\frac{\partial a^2}{\partial \xi }.  \tag{12.3}
\end{equation}
It follows from Eq. (12.2) that the azimuthal momentum is conserved, $%
p_\theta =const$, $\beta _\theta (\tau )=p_\theta (0)/\gamma (\tau )$. The
azimuthal momentum has not essential influence on the axial and radial
dynamics, and we suppose $p_\theta (0)=0$ in this paper. For the energy of
electrons Eq. (11) gives the following equation:

\begin{equation}
\frac{d\gamma }{d\tau }=-\beta _g(\mathbf{\beta E})-\frac 1{4\gamma }\frac{%
\partial a^2}{\partial \xi }.  \tag{13}
\end{equation}
From Eqs. (12.4), (13) and (9) we obtain the integral of motion

\begin{equation}
\gamma -\beta _gp_z-\Phi (\xi ,r)=const,  \tag{14}
\end{equation}
which formally coincides with the one-dimensional integral of motion (3)
[14,26,27]. Electrons can be trapped in the region where wakefield is both
accelerating and focusing. For the scattered particles, from Eq. (14) one
has $p_r^2=(S+\beta _gp_z)^2-p_z^2-1$, here $S=[1+\mathbf{p}%
^2(0)]^{1/2}-\beta _gp_z(0)$. If an electron is initially non-relativistic ($%
|\mathbf{p}(0)|\ll 1$, $S\approx 1$), then $p_r\approx (2p_z)^{1/2}$ and $%
\tan \theta =p_r/p_z\approx [2/(\gamma -1)]^{1/2}$, where $\theta $ is the
angle between $z$-axis and final momentum of the scattered electron [28].

Taking into account Eq. (13), we rewrite Eqs. (12.1) and (12.3) in the form

\begin{equation}
\frac{d^2\xi }{d\tau ^2}+\frac 1\gamma \left[ (1-\beta _z^2)E_z-\beta
_z\beta _rE_r+\beta _rH_\theta \right] +\frac{(1-\beta _g\beta _z)}{4\beta
_g^2\gamma ^2}\frac{\partial a^2}{\partial \xi }=0,  \tag{15.1}
\end{equation}

\begin{equation}
\frac{d^2r}{d\tau ^2}+\frac 1\gamma \left[ (1-\beta _r^2)E_r-\beta _z\beta
_rE_z-\beta _zH_\theta \right] -\frac 1{4\beta _g\gamma ^2}\left( \beta _r%
\frac{\partial a^2}{\partial \xi }-\frac 1{\beta _g}\frac{\partial a^2}{%
\partial r}\right) =0.  \tag{15.2}
\end{equation}
The normalized components of velocity obey the formulae $\beta _z=\beta
_g(1+d\xi /d\tau )$ and $\beta _r=\beta _gdr/d\tau $. For particles trapped
in the wakefield, we suppose that during acceleration $\beta _z\approx 1$, $%
\beta _r^2\ll 1$ and $r<1$ (the numerical results presented below show that
this is the case). Then, from Eq. (15.1) one has

\begin{equation}
d^2\xi /d\tau ^2\approx E_z/\gamma ^3.  \tag{16}
\end{equation}
It follows from this equation that $d\gamma /d\tau \approx -E_z$ and $\gamma
\approx -\int E_zd\tau $. Thus, the longitudinal dynamics of accelerating
particles is approximately the same as in the one-dimensional case. The
radial motion of electrons, according to Eq. (15.2), obeys the equation

\begin{equation}
\frac{d^2r}{d\tau ^2}+\frac{|E_z|}\gamma \frac{dr}{d\tau }+\Omega ^2r\approx
0,  \tag{17}
\end{equation}
where $\Omega =(|\partial f_r/\partial r|/\gamma )^{1/2}$ is the betatron
frequency. Supposing that the value of $E_z$ is approximately conserved
during acceleration, we can write $\gamma \approx |E_z|(\tau -\tau _{tr})$.
In this case solution of Eq. (17) is

\begin{equation}
r=r(\tau _{tr})J_0[2(|\partial f_r/\partial r|(\tau -\tau
_{tr})/|E_z|)^{1/2}],  \tag{18}
\end{equation}
where $J_0$ is the Bessel function of zero order.

\begin{center}
\textbf{C. Results of test-particle simulations and discussion}
\end{center}

Motion of test electrons in the 2D wakefield presented in Fig. 7 was
investigated by numerical solution of Eqs. (15.1) and (15.2) for different
initial positions and momentums. Figure 9 shows the behavior of electrons
with zero initial transverse momentums and with different initial radial
positions. One can see that particles are trapped near the first
accelerating maximum in the wake. During the trapping, electrons concentrate
near the axis due to the focusing force $\beta _zH_\theta -E_r$. Sins the
longitudinal size of the trapped bunch is much less than the plasma
wavelength and its transverse size is essentially less than that of the
laser pulse, the electrons experience approximately the same accelerating
field; the longitudinal dynamics is well described by the one-dimensional
theory. The focusing force acting on the bunch electrons depends on $r$
linearly (see Fig. 8). The small bunch sizes (as compared with the wakefield
characteristic sizes) and the fact that electrons are trapped near the
accelerating maximum provide high accelerating gradient and relatively small
energy spread. For example, the relative energy spread of electrons
presented in Fig. 9 is $5\%$ at $\tau =300$. The numerical results show that
dynamics of the accelerating bunch is well described by approximate
equations (16)-(18). The betatron oscillations of the accelerating electrons
are clearly seen in Fig. 9(b). The wavelength of this oscillations decreases
with the increase of particle's energy that conform to the formula for
betatron frequency. Radial velocity of accelerating electrons is much less
than the longitudinal one, $|\beta _r(\tau )|<0.1$. One can see also that
even electrons which are initially at the periphery [$r(\tau =0)\equiv
r_0\sim \sigma _r$] can be trapped in the wakefield and accelerated. The
characteristic dependence of the minimum trapping threshold $p_{z,\min }$ on
the initial radial position of an electron is presented in Fig. 10. Figure
11 shows the minimum and maximum initial radial momentums of trapped
electrons in dependence on initial radial position. The figure witnesses
that electrons which initially move at relatively high angle to the axis (up
to tens degrees) also can be trapped and accelerated. This again is caused
by the focusing force and the fact that electrons are initially
non-relativistic ($\gamma _0\sim 1$).

In Fig. 12 we show behavior of electrons of a bunch with the following
initial parameters: $0\leq \xi _0\leq 5$, $r_0\leq 4$, $0.6\leq p_{z0}\leq
0.8$, $-0.02\leq p_{r0}\leq 0.02$. The passing particles (not showed) are
well separated from accelerating one both spatially and energetically. The
length of accelerating bunch in this case also is much less than the plasma
wavelength [$L(\tau =100)\approx 0.27$, $L(\tau =300)\approx 0.19$]. The
radius of the bunch $R$ decreases relatively slowly during acceleration and
is essentially less than the characteristic transverse size of the wakefield 
$\sigma _r$, $R(\tau )\sim 1$; the bunch radius can be reduced by the choice
of smaller laser spot size. The absolute energy spread does not change
practically, $\Delta \gamma \approx 24$, but the relative energy spread
falls and is equal to about 10\% at $\tau =300$. The estimations of absolute
and relative energy spreads presented in Sec. II are valid also in 3D case.

Total number of electrons trapped and their density can be estimated from
expressions

\begin{equation}
N_{tot}\sim \delta \pi n_{b0}\sigma _r^2L_0/k_p^3,  \tag{19}
\end{equation}

\begin{equation}
n_b\sim \delta n_{b0}(\sigma _r/R)^2(L_0/L).  \tag{20}
\end{equation}
The on-axis amplitude of the linear wake wave excited by the bunch is
reduced by the factor $T(R)=1-RK_1(R)<1$ [29] (where $K_1$ is the modified
Bessel function) as compared to the one-dimensional case (see Sec. II).
Therefore, in our case, for the amplitude of wakefield generated by the
bunch, we have: $E_{b,\max }\approx TLn_b/n_p$. This wakefield can be
neglected when $E_{b,\max }\ll E_{z,\max }$, or taking into account (20) -
if $\delta TL_0(\sigma _r/R)^2(n_{b0}/n_p)\ll 1$; when $R\ll 1$, $T\approx
R^2/2$, and this condition reads $\delta L_0\sigma _r^2n_{b0}/2n_p\ll 1$.
Total number of bunch electrons, according to (19), is restricted by the
following condition: $N_{tot}\ll \pi n_pk_p^{-3}(R^2/T)\approx 1.4\times
10^7(R^2/T)\lambda _p[\mu m]$.

For the normalized emittance $\varepsilon _n=\sigma _0^2/\beta $ (here $%
\sigma _0$ is the matched transverse size of the bunch, $\beta $ is the
betatron length) of the accelerating bunch, in our notations, one can write $%
\varepsilon _n\sim R^2\Omega \lambda _p/4\pi ^2$. In the case $\lambda
_p=100\mu m$ ($n_p\approx 10^{17}cm^{-3}$), for the bunch presented in Fig.
12, $\varepsilon _n\sim 8nm/\gamma ^{1/2}$; for example, $\varepsilon _n\sim
0.5nm$ when $\gamma =300$, that is comparable with the emittance expected in
the TeV-range laser wakefield accelerator [30,31] (see also Refs. [14,31-34]
for the dynamics of accelerating bunch).

\begin{center}
\textbf{IV. SUMMARY}
\end{center}

The results of the present work show the possibility of trapping, essential
compression both in longitudinal and transverse directions and
ultra-relativistic acceleration of an initially non-relativistic or
weakly-relativistic electron bunch in moderately nonlinear ($a_0\sim 1$, $%
E_{z,\max }\sim 1$) laser wakefield. The initial bunch can be generated, for
example, by a photocathode. So far as electron bunch is initially
non-relativistic ($\gamma _0\sim 1$), trapping and compression take place
during time interval comparable with the plasma wave period, that is much
less than the time scale of longitudinal dynamics of relativistic particles
in the wake [33]. Due to the fact that trapped bunch sizes are essentially
less than characteristic spatial scales of the wake wave, the energy spread
in the accelerated bunch can be relatively low, namely few percent. In our
scheme the problems connected with the wake wavefront curvature also are
removed. The electron bunch trapped and accelerated can be accelerated
further in the multi-stage LWFA [31].

Thus, the scheme of LWFA proposed, has the following advantages: (a) instead
of injection of an enough dense relativistic electron bunch with small sizes
(in order of a micrometer), our scheme utilizes a non-relativistic, rare and
long electron bunch, that is much easier to get technically, (b) femtosecond
electron bunch synchronization in the laser wakefield is not required, (c)
effective electron bunch compression, and (d) spatial and energetic
separation of the initial electron bunch, that can decrease the trapped
bunch emittance.

\begin{center}
\textbf{ACKNOWLEDGMENT}
\end{center}

The author is grateful to B. Hafizi, R. Hubbard and P. Sprangle (Naval
Research Laboratory, Washington, DC) for helpful discussions.

\begin{center}
\textbf{REFERENCES}
\end{center}

[1] D. Strickland and G. Mourou, Opt. Commun. \textbf{56}, 219 (1985).

[2] G. A. Mourou, C. P. Barty, and M. D. Perry, Phys. Today \textbf{51}, 22
(1998).

[3] $\,$E. Esarey, P. Sprangle, J. Krall, and A. Ting, IEEE Trans. Plasma
Sci. \textbf{24}, 252 (1996).

[4] P. Sprangle, E. Esarey, and A. Ting, Phys. Rev. A \textbf{41}, 4463
(1990).

[5] N. E. Burnett and P. B. Corkum, J. Opt. Soc. Am. B \textbf{6}, 1195
(1989).

[6] M. Tabak, J. Hammer, M. E. Glinsky, W. L. Kruer, S. C. Wilks, J.
Woodworth, E. M. Campbell, M. D. Perry, and R. J. Mason , Phys. Plasmas 
\textbf{1}, 1626 (1994).

[7] T. Tajima and J. M. Dawson, Phys. Rev. Lett. \textbf{43}, 267 (1979).

[8] $\,$L. M. Gorbunov and V. I. Kirsanov, Zh. Eksp. Teor. Fiz. \textbf{93},
509 (1987) [Sov. Phys. JETP \textbf{66}, 290 (1987)].

[9] C. I. Moore, A. Ting, S. J. McNaught, J. Qiu, H. R. Burris, and P.
Sprangle, Phys. Rev. Lett. \textbf{82}, 1688 (1999).

[10] D. Umstadter, J. K. Kim, and E. Dodd, Phys. Rev. Lett. \textbf{76},
2073 (1996).

[11] E. Esarey, R. F. Hubbard, W. P. Leemans, A. Ting, and P. Sprangle,
Phys. Rev. Lett. \textbf{79}, 2682 (1997); E. Esarey, C. B. Schroder, W. P.
Leemans, and B. Hafizi, Phys. Plasmas \textbf{6}, 2262 (1999).

[12] $\,$E. Esarey, P. Sprangle, J. Krall, and A. Ting, IEEE J. Quantum
Electron. \textbf{33}, 1879 (1997).

[13] N. E. Andreev, L. M. Gorbunov, V. I. Kirsanov, K. Nakajima, and A.
Ogata, Phys. Plasmas \textbf{4}, 1145 (1997).

[14] A. J. W. Reitsma, V. V. Goloviznin, L. P. J. Kamp, and T. J. Schep,
Phys. Rev. E \textbf{63}, 046502 (2001).

[15] A. G. Khachatryan, Phys. Rev. E \textbf{60}, 6210 (1999).

[16] A. G. Khachatryan, Phys. Plasmas \textbf{7}, 5252 (2000).

[17] G. Shvets, N. J. Fisch, and A. Pukhov, IEEE Trans. Plasma Sci. \textbf{%
28}, 1194 (2000).

[18] C. Du and Z. Xu, Phys. Plasmas \textbf{7}, 1582 (2000).

[19] \thinspace A. I. Akhiezer and R .V. Polovin, Zh. Eksp. Teor. Fiz. 
\textbf{30}, 915 (1956) [Sov. Phys. JETP \textbf{3}, 696 (1956)].

[20] A. G. Khachatryan, Phys. Plasmas \textbf{4}, 4136 (1997).

[21] E. Esarey and M. Pilloff, Phys. Plasmas \textbf{2}, 1432 (1995).

[22] \thinspace R. D. Ruth, A. W. Chao, P. L. Morton, and P. B. Wilson,
Part. Accel. \textbf{17}, 171 (1985).

[23] A. J. Mackinnon, M. Borghesi, A. Iwase, and O. Willi, Phys. Rev. Lett. 
\textbf{80}, 5349 (1998); E. De Wispelaere, V. Malka, S. H\"uller, F.
Amiranoff, S. Baton, R. Bonadio, M. Casanova, F. Dorchies, R. Haroutunian,
and A. Modena, Phys. Rev. E \textbf{59}, 7110 (1999); G. S. Sarkisov, V. Yu.
Bychenkov, V. N. Novikov, V. T. Tikhonchuk, A. Maksimchuk, S.-Y. Chen, R.
Wagner, G. Mourou, and D. Umstadter, \textit{ibid}. \textbf{59}, 7042
(1999); T. R. Clark and H. M. Michberg, Phys. Plasmas \textbf{7}, 2192
(2000); J. Faure, V. Malka, J.-R. Marqu\`es, F. Amiranoff, C. Courtois, Z.
Najmudin, K. Krushelnick, M. Salvati, A. E. Dangor, A. Solodov, P. Mora,
J.-C. Adam, and A. H\'eron, \textit{ibid}. \textbf{7}, 3009 (2000); E. W.
Gaul, S. P. Le Blanc, A. R. Rundquist, R. Zgadzaj, H. Langhoff, and M. C.
Downer, Applied Phys. Lett. \textbf{77}, 4112 (2000).

[24] A. G. Khachatryan, Fizika Plazmy \textbf{27}, 921 (2001) [Plasma Phys.
Rep. \textbf{27}, 860 (2001)].

[25] S. V. Bulanov, F. Pegoraro, A. M. Pukhov, and A. S. Sakharov, Phys.
Rev. Lett. \textbf{78}, 4205 (1997); S. V. Bulanov, F. Pegoraro, and J.
Sakai, Nucl. Instrum. Methods Phys. Res. A \textbf{410}, 477 (1998).

[26] P. Mora and T. M. Antonsen, Phys. Rev. E \textbf{53}, R2068 (1996).

[27] S.-Y. Chen, M. Krishnan, A. Maksimchuk, and D. Umstadter, Phys. Plasmas 
\textbf{6}, 4739 (1999).

[28] P. B. Corcum, N. H. Burnett, and F. Brunel, in \textit{Atoms in Intense
Fields}, edited by M. Gavrila (Academic Press, New York, 1992).

[29] R. Keinigs and M. Jones, Phys. Fluids \textbf{30}, 252 (1987); A. G.
Khachatryan, A. Ts. Amatuni, E. V. Sekhposyan, and S. S. Elbakyan, Fizika
Plazmy \textbf{22}, 638 (1996) [Plasma Phys. Rep. \textbf{22}, 576 (1996)].

[30] M. Xie, T. Tajima, K. Yokoya, and S. Chattopadhyay, in \textit{Advanced
Accelerator Concepts: Seventh Workshop, Lake Tahoe, 1996}, edited by S.
Chattopadhyay (AIP, New York, 1997), p. 233.

[31] S. Cheshkov, T. Tajima, and W. Horton, Phys. Rev. ST-AB \textbf{3},
071301 (2000).

[32] R. Assman and K. Yokoya, Nucl. Instrum. Methods Phys. Res. A \textbf{410%
}, 544 (1998).

[33] N. E. Andreev, S. V. Kuznetsov, and I. V. Pogorelsky, Phys. Rev. ST-AB 
\textbf{3}, 021301 (2000).

[34] C. Chiu, S. Cheshkov, and T. Tajima, Phys. Rev. ST-AB \textbf{3},
101301 (2000).

\newpage\ 

\begin{center}
\textbf{FIGURE CAPTIONS}
\end{center}

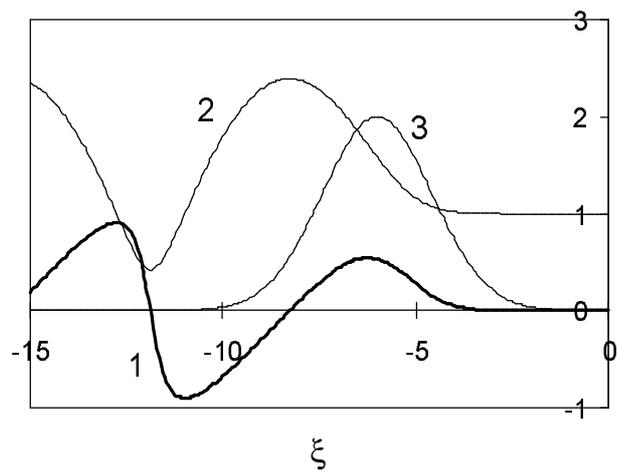
Fig. 1. The one-dimensional nonlinear wakefield excited by the
linearly-polarized laser pulse with peak normalized amplitude $a_0=2,$ $%
\sigma _z=2$, $\gamma =10$. 1 - The normalized longitudinal electric field $%
E_z(\xi )$; 2 - the dimensionless potential of the wakefield $\Phi (\xi )$;
3 - the amplitude of the laser pulse $a(\xi )$.

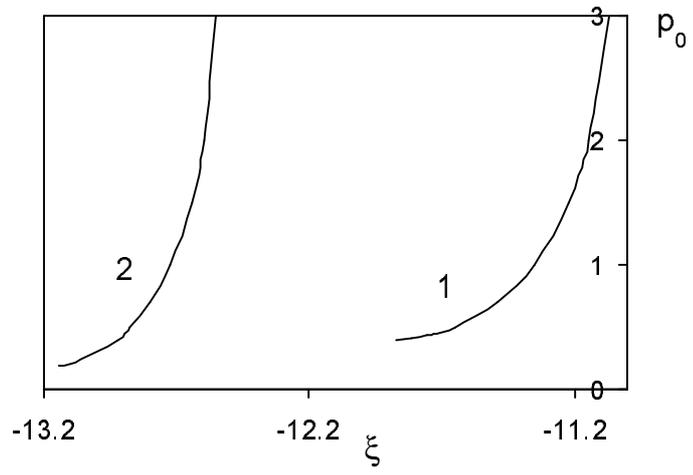
Fig. 2. Dependence of electron initial momentum $p_0$ on the trapping point
near the first accelerating maximum. 1 - $a_0=2$; 2 - $a_0=3$.

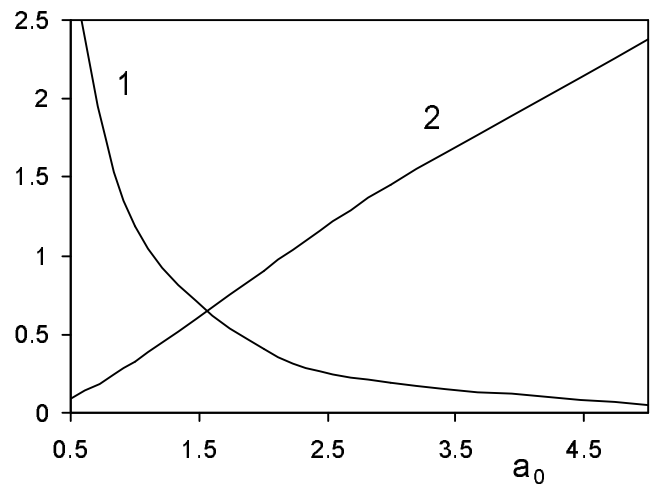
Fig. 3. The minimum momentum of the trapped electrons $p_{\min }$ (curve 1)
and the wake wave amplitude $E_{z,\max }$ (curve 2) in dependence on peak
amplitude of the laser pulse $a_0$.

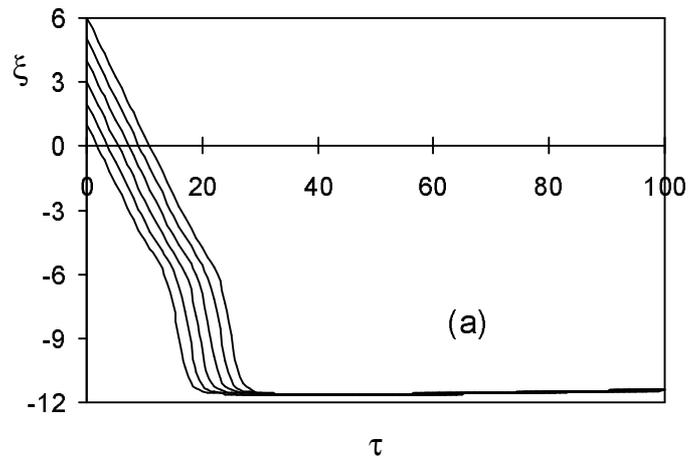
Fig. 4. Trapping, compression and acceleration of initially mono-energetic
electron bunch in the wakefield presented in Fig. 1, $p_0=0.5$, $1\leq \xi
_0\leq 6$. Evolution of the coordinate (a) and the relativistic factor (b)
of electrons.

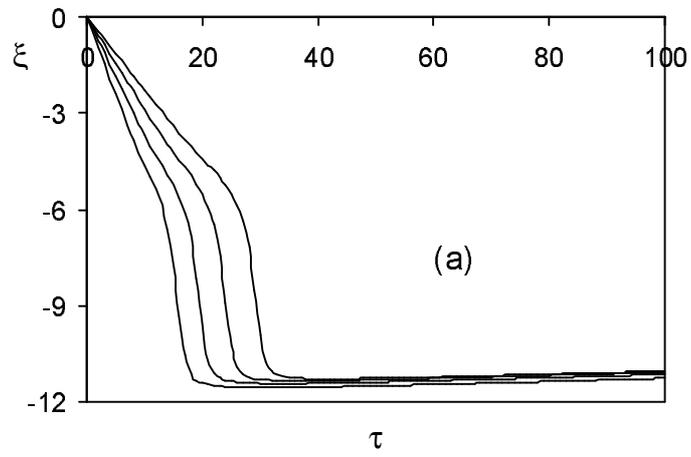
Fig. 5. Behavior of electrons with initial position $\xi _0=0$ and with
initial momentums $p_0=0.5,$ $0.8,$ $1$ and $1.2$ in the wakefield shown in
Fig. 1. Electrons with smaller initial momentums are trapped earlier. (a)
Coordinate and (b) relativistic factor of electrons.

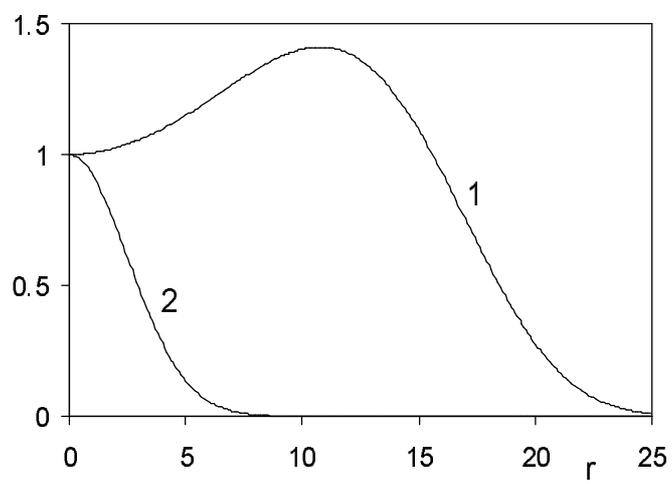
Fig. 6. The radial profiles of unperturbed electron density in the plasma
channel (curve 1) and the laser pulse (curve 2), $r_{ch}=\sigma _r=5$, $%
b=0.01$.

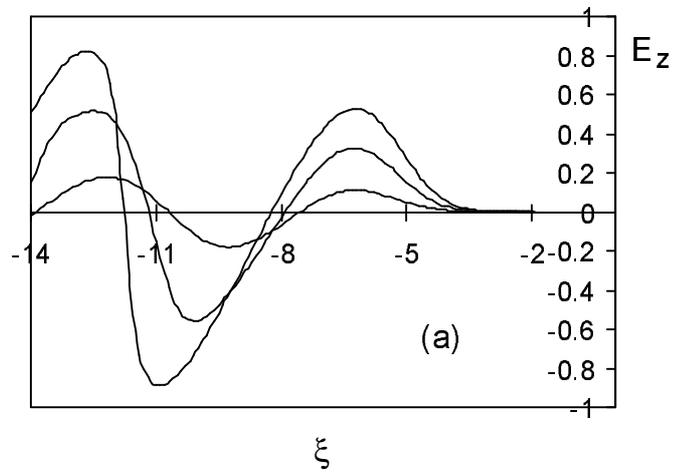
Fig. 7. The two-dimensional nonlinear laser wakefield excited in the plasma
channel with the radial density profile shown in Fig. 6, $a_0=2$, $\sigma
_z=2$, $\sigma _r=5$. (a) The longitudinal electric field for $r_0=0$, $3$
and $5$ in the order of magnitude reduction. (b) The focusing field $%
f_r=\beta _gH_\theta -E_r$. 1 - $r=1$; 2 - $r=3$; 3 - $r=5$.

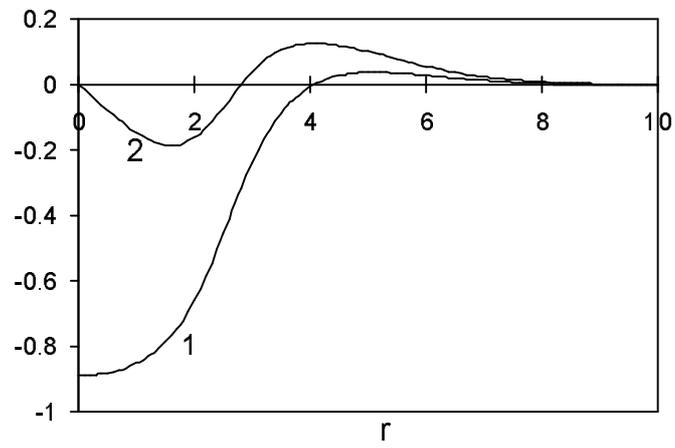
Fig. 8. The radial behavior of the wakefield shown in Fig. 7, at $\xi =-10.9$%
. 1 - longitudinal electric field $E_z(\xi =-10.9,r)$; 2 - the focusing
force $f_r(\xi =-10.9,r)$.

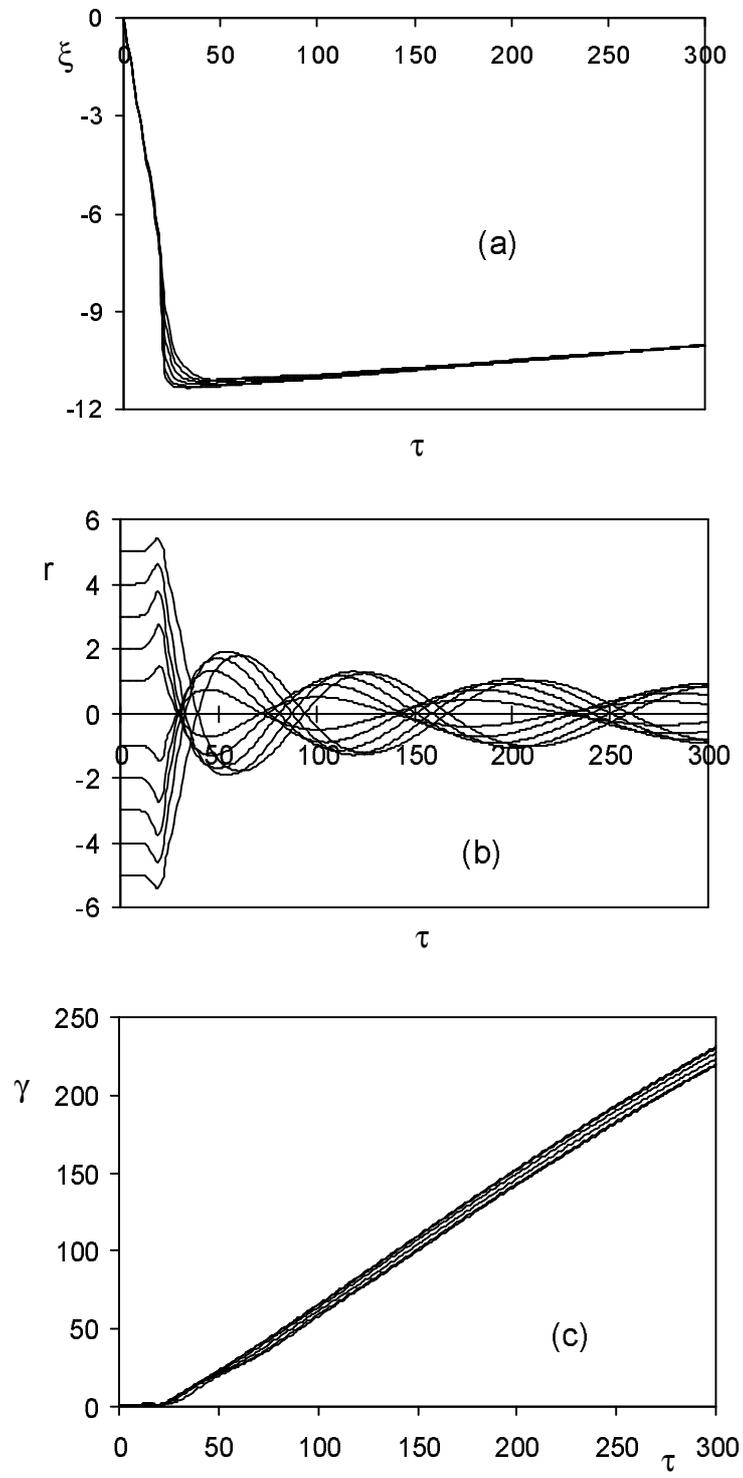
Fig. 9. Trapping and acceleration of electrons with zero initial momentums
in the wakefield given in Fig. 7, $p_{z0}=0.8$, $\xi _0=0$. Longitudinal (a)
and radial (b) positions and relativistic factor (c) of electrons.

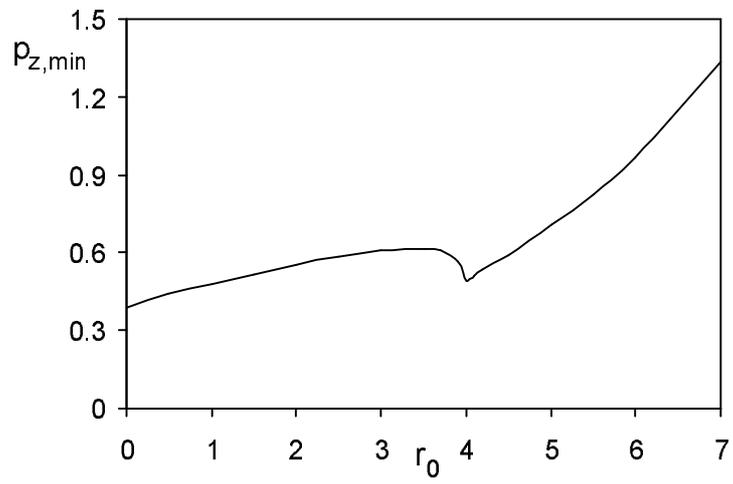
Fig. 10. The characteristic dependence of the minimum trapping threshold on
initial radial position of electron, $p_{r0}=0$, $\xi _0=0$.

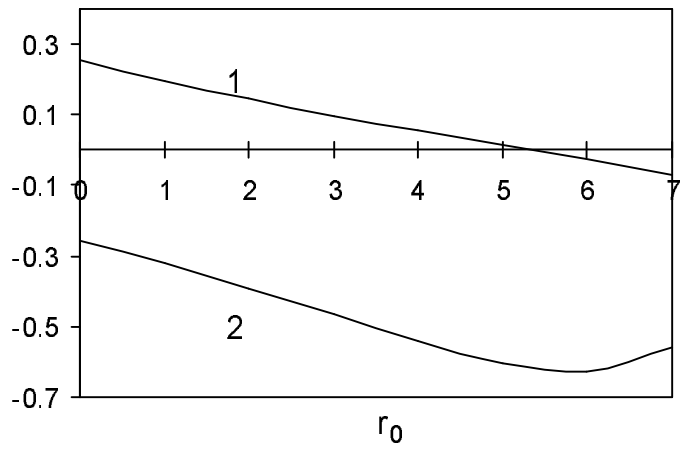
Fig. 11. Maximum (curve 1) and minimum (curve 2) initial radial momentums of
trapped electrons depending on initial radial position, $p_{z0}=0.8$, $\xi
_0=0$.

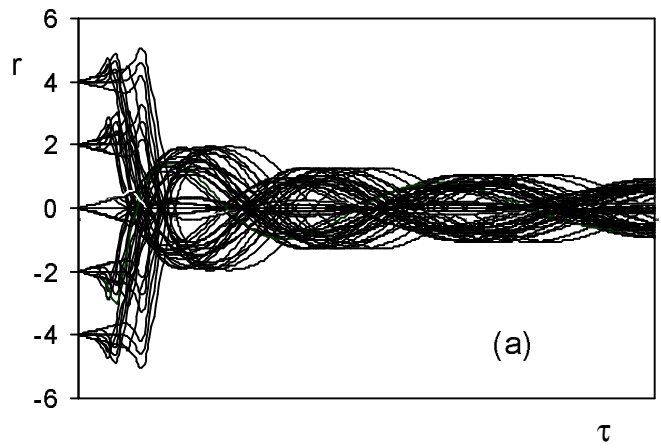
Fig. 12. Trapping, compression and acceleration of an electron bunch in the
wakefield presented in Fig. 7. Initial parameters of the bunch are: $0\leq
\xi _0\leq 5$, $r_0\leq 4$, $0.6\leq p_{z0}\leq 0.8$, $-0.02\leq p_{r0}\leq
0.02$. Radial positions (a) and relativistic factor (b) of electrons.

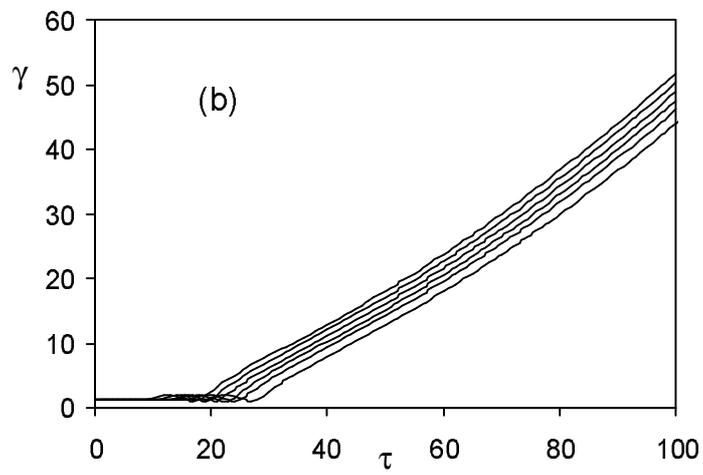
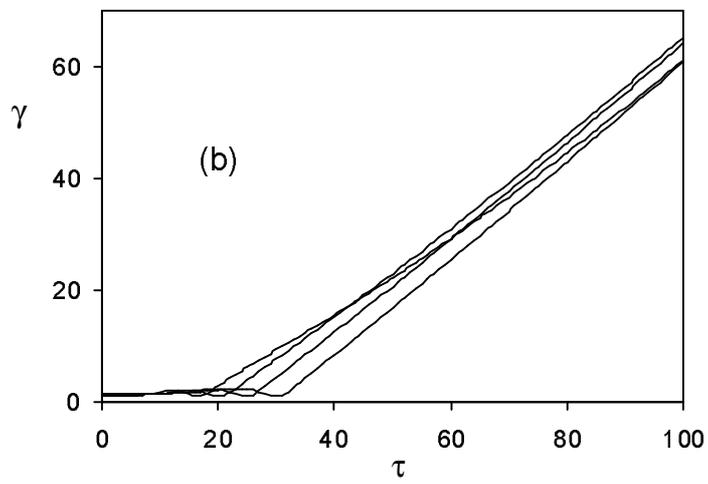
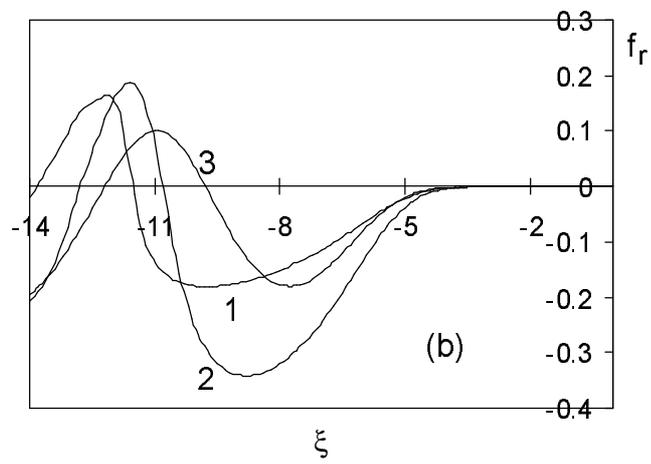
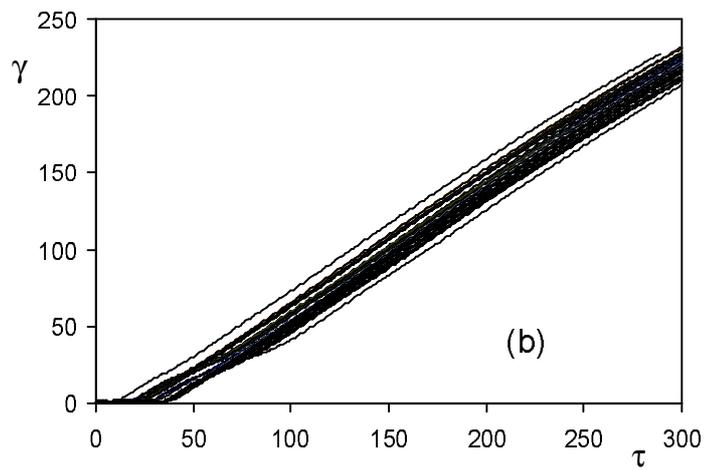
\end{document}